\documentclass[aps,twocolumns,eqsecnum,preprintnumbers,showpacs,amsmath,amssymb]
{revtex4}

\usepackage{graphicx}
\usepackage{dcolumn}
\usepackage{bm}

\begin{document}

\title{ GLUON CONFINEMENT CRITERION IN QCD }

\author{V. Gogohia}
\email[]{gogohia@rmki.kfki.hu}

\affiliation{HAS, CRIP, RMKI, Depart. Theor. Phys., Budapest 114,
P.O.B. 49, H-1525, Hungary}

\date{\today}

\begin{abstract}
We fix exactly and uniquely the infrared structure of the full
gluon propagator in  QCD, not solving explicitly the corresponding
dynamical equation of motion. By construction, this structure is
an infinite sum over all possible severe (i.e., more singular than
$1/ q^2$) infrared singularities. It reflects the zero momentum
modes enhancement effect in the true QCD vacuum, which is due to
the self-interaction of massless gluons. Its existence
automatically exhibits a characteristic mass (the so-called mass
gap). It is responsible for the scale of nonperturbative dynamics
in the true QCD ground state. The theory of distributions,
complemented by the dimensional regularization method, allows one
to put the severe infrared singularities under firm mathematical
control. By an infrared renormalization of a mass gap only, the
infrared structure of the full gluon propagator is exactly reduced
to the simplest severe infrared singularity, the famous
$(q^2)^{-2}$. Thus we have exactly established the interaction
between quarks (concerning its pure gluon (i.e., nonlinear)
contribution) up to its unimportant perturbative part. This also
makes it possible for the first time to formulate the gluon
confinement criterion and intrinsically nonperturbative phase in
QCD in a manifestly gauge-invariant ways.
\end{abstract}



\maketitle

\section{Introduction}

To say today that QCD is a nonperturbative (NP) theory is almost a
tautology. The problem is how to define it exactly, since we know
for sure that QCD has a perturbative (PT) phase as well because of
asymptotic freedom (AF) \cite{1}. In order to define exactly the
NP phase in QCD, let us start with one of the main objects in the
Yang-Mills (YM) sector. The two-point Green's function, describing
the full gluon propagator, is (using Euclidean signature here and
everywhere below)

\begin{equation}
D_{\mu\nu}(q) = i \left\{ T_{\mu\nu}(q)d(q^2, \xi) + \xi
L_{\mu\nu}(q) \right\} {1 \over q^2 },
\end{equation}
where $\xi$  is the gauge fixing parameter ($\xi = 0$ -- Landau
gauge, $\xi = 1$ -- Feynman gauge) and
$T_{\mu\nu}(q)=g_{\mu\nu}-(q_{\mu} q_{\nu} / q^2) = g_{\mu\nu } -
L_{\mu\nu}(q)$. Evidently, $T_{\mu\nu}(q)$ is the transverse
(physical) component of the full gluon propagator, while
$L_{\mu\nu}(q)$ is its longitudinal (unphysical) one. The free
gluon propagator is obtained by simply setting the full gluon form
factor $d(q^2, \xi)=1$ in Eq. (1.1). The dynamical equation of
motion for the full gluon propagator is the so-called gluon
Schwinger-Dyson (SD) equation, which is part of the whole SD
system of dynamical equations of motion \cite{1}. The solutions of
the gluon SD equation are supposed to reflect the complexity of
the quantum structure of the QCD ground state. Precisely this
determines one of the central roles of the full gluon propagator
in the SD system of equations. The SD equation for the full gluon
propagator is a highly nonlinear system of four-dimensional
integrals, containing many different, unknown in general,
propagators and vertices, which, in their turn, satisfy too
complicated integral equations, containing different scattering
amplitudes and kernels, so there is no hope for exact solution(s).
However, in any case the solutions of this equation can be
distinguished from each other by their behavior in the infrared
(IR) limit, describing thus many (several) different types of
quantum excitations and fluctuations of gluon field configurations
in the QCD vacuum. The ultraviolet (UV) limit of these solutions
is uniquely fixed by AF.

The IR asymptotics of the full gluon propagator can be either
singular or smooth. However, the smooth behavior of the full gluon
propagator (1.1) is possible only in one exceptional covariant
gauge - the Landau gauge ($\xi = 0$) \cite{2}, i.e., it is a gauge
artifact solution in this case. Being thus a gauge artifact, it
can be related to none of the physical phenomena such as quark and
gluon confinement or dynamical breakdown of chiral symmetry
(DBCS), which are, by definition, manifestly gauge-invariant. To
our best knowledge, beyond the covariant gauges, other than the
Landau gauge, the smooth behavior is not known. Anyway, nobody
knows how to relate the smooth asymptotics in any covariant gauge
to color confinement phenomenon, DBCS, etc. For example, it does
not provide a linearly rising potential between heavy quarks
"seen" by lattice QCD simulations \cite{3}. Hence we will not
discuss it in what follows, though a solution with smooth
asymptotics may exist as a formal one to the gluon SD equation.

Thus we are left with the IR singular behavior of the full gluon
propagator only, which is possible in any gauge (in principle, the
free gluon propagator can be also used in any gauge. The Feynman
gauge free gluon propagator in the IR has been used by Gribov
\cite{4} in order to investigate quark confinement within
precisely the SD system of equations approach). The only problem
is to decide which type of the IR singularities is to be accounted
for. The free gluon propagator (see Eq. (1.1) with $d(q^2)=1$) has
an exact power-type $1/q^2$ IR singularity. So the IR
singularities as much singular as $1/q^2$ as $q^2 \rightarrow 0$
will be called PT IR singularities. The IR singularities which are
more severe than the above-mentioned exact power-type IR
singularity of the free gluon propagator will be called NP IR
singularities, i.e., they are more severe than $1/q^2$ as $q^2
\rightarrow 0$. They should be summarized (accumulated) into the
full gluon propagator and described effectively correctly by its
structure in the deep IR domain. Let us remind that for a long
time from the very beginning of QCD it has been already well known
that the QCD vacuum is really beset with the severe (or
equivalently NP) IR singularities if standard PT is applied
\cite{1,5,6,7,8,9,10,11,12}. "But it is to just this violent IR
behavior that we must look for the key to the low energy and large
distance hadron phenomena. In particular, the absence of quarks
and other colored objects can only be understood in terms of the
IR divergences in the self-energy of a color bearing objects"
\cite{10}. It is worth emphasizing that in Ref. \cite{13} it is
explicitly shown how the severe IR singularities inevitably appear
in the QCD vacuum, providing thus the basis for the zero momentum
modes enhancement (ZMME) effect there. So it is intrinsically
peculiar to the true QCD ground state due to the self-interaction
of massless gluons. Precisely this effect is reflected by the
appearance of the severe IR singularities in the gluon propagator.

It is clear also that any deviation of the full gluon propagator
from the free one in the IR, automatically requires an
introduction of the corresponding mass scale parameter,
responsible for the nontrivial dynamics in the IR region, the
so-called mass gap (see below). This is important, since there is
none explicitly present in the QCD Lagrangian (current quark mass
cannot be considered as a mass gap, since it is not the
renormalization group invariant). Of course, such gluon field
configurations, which are to be described by the severely IR
structure of the full gluon propagator, can be only of dynamical
origin. The only dynamical mechanism in QCD which can produce such
configurations in the vacuum, is the self-interaction of massless
gluons. So the above-mentioned mass gap appears on dynamical
ground. Let us remind that precisely this self-interaction in the
UV limit leads to AF.

The main purpose of this Letter is to establish exactly the deep
IR structure of the full gluon propagator, not solving the gluon
SD equation directly, which is a formidable task, anyway. On this
basis we will be able to derive the gluon confinement criterion in
a manifestly gauge-invariant way. However, it is convenient first
to emphasize the distribution nature of the severe (i.e., NP) IR
singularities. For this purpose, in the next section we will
explicitly introduce a few useful formulae from the distribution
theory (DT) \cite{14}, complemented by the dimensional
regularization (DR) method \cite{15}, which are crucial in our
investigation.

\section{ IR dimensional regularization within the distribution theory}

In general, all the Green's functions in QCD are generalized
functions, i.e., they are distributions. This is true especially
for the NP IR singularities due to the self-interaction of
massless gluons in the QCD vacuum. They present a rather broad and
important class of functions with algebraic singularities, i.e.,
functions with nonsummable singularities at isolated points \cite
{14} (at zero in our case). Roughly speaking, this means that all
relations involving distributions should be considered under
corresponding integrals, taking into account the smoothness
properties of the corresponding class of test functions (for
example, $\varphi(q)$ below). In principle, any regularization
scheme (i.e., how to parameterize the severe IR singularities and
thereby to put them under control) can be used; it should,
however, be compatible with DT \cite{14}.

Let us consider the positive definite ($P>0$) squared (quadratic)
Euclidean form $P(q) = q_0^2 +  q_1^2 + q_2^2 + ... + q_{n-1}^2 =
q^2$, where $n$ is the number of the components. The generalized
function (distribution) $P^{\lambda}(q)$, where $\lambda$ is, in
general, an arbitrary complex number, is defined as $(P^{\lambda},
\varphi) = \int_{P>0}P^{\lambda}(q) \varphi(q) d^nq$. At $Re
\lambda \geq 0$ this integral is convergent and is an analytic
function of $\lambda$. Analytical continuation to the region $Re
\lambda < 0$ shows that it has a simple pole at points \cite{14}

\begin{equation}
\lambda = - {n \over 2} - k, \quad k=0, 1, 2 ,3...
\end{equation}

In order to actually define the system of the SD equations in the
deep IR domain, it is necessary to introduce the IR regularization
parameter $\epsilon$, defined as $D = n + 2 \epsilon, \ \epsilon
\rightarrow 0^+$ within a gauge-invariant DR method \cite{15}. As
a result, all the Green's functions and "bare" parameters should
be regularized with respect to $\epsilon$ (see next sections),
which is to be set to zero at the end of the computations. The
structure of the NP IR singularities is then determined (when $n$
is even number) as follows \cite{14}:

\begin{equation}
(q^2)^{\lambda} = { C_{-1}^{(k)} \over \lambda +(D/2) + k} +
finite \ terms,
\end{equation}
where the residue is

\begin{equation}
 C_{-1}^{(k)} = { \pi^{n/2} \over 2^{2k} k! \Gamma ((n/2) + k) } \times
L^k \delta^n (q)
\end{equation}
with $L = (\partial^2 / \partial q^2_0) + (\partial^2 /
\partial q^2_1) + ... + (\partial^2 / \partial q^2_{n-1})$.

Thus the regularization of the NP IR singularities (2.2) is
nothing but the so-called Laurent expansion that is dimensionally
regularized. Let us underline its most remarkable feature. The
order of singularity does not depend on $\lambda$, $n$ and $k$. In
terms of the IR regularization parameter $\epsilon$, it is always
a simple pole $1/ \epsilon$. This means that all power terms in
Eq. (2.2) will have the same singularity, i.e.,

\begin{equation}
(q^2)^{- {n \over 2} - k } = { 1 \over \epsilon} C_{-1}^{(k)} +
finite \ terms, \quad \epsilon \rightarrow 0^+,
\end{equation}
where we can put $D=n$ now (i.e., after introducing this
expansion). By "$finite \ terms$" here and everywhere a number of
necessary subtractions under corresponding integrals is
understood. However, the residue at a pole will be drastically
changed from one power singularity to another. This means
different solutions to the whole system of the SD equations for
different set of numbers $\lambda$ and $k$. Different solutions
mean, in turn, different vacua. Thus in this picture different
vacua are to be labelled by two independent numbers: the exponent
$\lambda$ and $k$. At given number of $D(=n)$ the exponent
$\lambda$ is always negative being integer if $D(=n)$ is an even
number or fractional if $D(=n)$ is an odd number. The number $k$
is always integer and positive and precisely it determines the
corresponding residue at the pole, see Eq. (2.3). It would be not
surprising if these numbers were somehow related to the nontrivial
topology of the nD QCD vacuum.

Concluding, let us note that the structure of the severe IR
singularities in Euclidean space is much simpler than in Minkowski
space, where kinematical (unphysical) singularities due to the
light cone also exist \cite{1,14}. In this case it is rather
difficult to untangle them correctly from the dynamical
singularities, the only ones which are important for the
calculation of any physical observable. Also the consideration is
much more complicated in configuration space \cite{14}. That is
why we always prefer to work in momentum space (where propagators
do not depend explicitly on the number of dimensions) with
Euclidean signature. We also prefer to work in covariant gauges in
order to avoid peculiarities of the noncovariant gauges
\cite{1,16}, for example how to untangle the gauge pole from the
dynamical one.

\section{The general structure of the full gluon propagator}

 For the above-mentioned purpose, namely, how to define the NP phase in QCD,
 it is convenient to begin with the exact decomposition of the full gluon form
factor in Eq. (1.1) as follows:

\begin{equation}
d(q^2) = d(q^2) - d^{PT}(q^2) + d^{PT}(q^2) = d^{NP}(q^2) +
d^{PT}(q^2),
\end{equation}
where, for simplicity, the dependence on the gauge fixing
parameter is omitted. In fact, this formal equation represents one
unknown function (the full gluon form factor) as an exact sum of
the two other unknown functions, which can be always done. So at
this stage there is no approximation made (only exact algebraic
manipulations). We would like to let the PT part of this exact
decomposition to be responsible for the known UV asymptotics
(since it is fixed by AF) of the full gluon propagator, while the
NP part is chosen to be responsible for its unknown yet IR
asymptotics. It is worth emphasizing that in realistic models of
the full gluon propagator, the NP part usually correctly
reproduces its deep IR asymptotics, determining thus the strong
intrinsic influence of the IR properties of the theory on its NP
dynamics. Evidently, the decomposition (3.1) represents an exact
subtraction of the PT contribution at the fundamental gluon level,
and consequently both terms in the right-hand-side of Eq. (3.1)
are determined in the whole momentum range, $[0, \infty)$. Let us
emphasize that the exact gluon form factor $d(q^2)$ being also NP,
nevertheless, is "contaminated" by the PT contributions, while
$d^{NP}(q^2)$ due to the subtraction (3.1) is free of them, i.e.,
it is truly NP.

Substituting the exact decomposition (3.1) into the full gluon
propagator (1.1), one obtains

\begin{equation}
D_{\mu\nu}(q) = D^{INP}_{\mu\nu}(q) + D^{PT}_{\mu\nu}(q),
\end{equation}
where

\begin{equation}
 D^{INP}_{\mu\nu}(q) = i T_{\mu\nu}(q) d^{NP}(q^2){ 1 \over q^2}=
i T_{\mu\nu}(q) d^{INP}(q^2),
\end{equation}

\begin{equation}
D^{PT}_{\mu\nu}(q) = i \{ T_{\mu\nu}(q)  d^{PT}(q^2) + \xi
L_{\mu\nu}(q) \} { 1 \over q^2}.
\end{equation}
Here the superscript "INP" is the short-hand notation for
intrinsically NP. Its definition will given below. The exact
decomposition (3.2) has a remarkable feature.
 The explicit gauge dependence of the
full gluon propagator is exactly shifted from its INP part to its
PT part. In other words, we want the INP part to be always
transverse, while leaving the PT part to be of arbitrary gauge.
This exact separation will have also a dynamical ground. It is
clear also that the PT part of the full gluon propagator is, by
definition, as much singular as the free gluon propagator's exact
power-type IR singularity. This is the first reason why the
longitudinal part of the full gluon propagator, which has the same
exact IR singularity, has been shifted to its PT part, and the
existence of which is determined by AF.

 As was mentioned above, we want the INP gluon form factor
$d^{INP}(q^2)$ to be responsible for the deep IR structure of the
full gluon propagator, which is saturated by the severe IR
singularities. So there is a problem how to take them into account
analytically in terms of the full gluon propagator. For this aim,
it is convenient to introduce the auxiliary INP gluon form factor
as follows:

\begin{equation}
d^{INP}_{\lambda_k}(q^2,
 \Delta^2)= (\Delta^2)^{-\lambda_k - 1} (q^2)^{\lambda_k}
 f_{\lambda_k}(q^2),
\end{equation}
where the exponent $\lambda_k$ is, in general, an arbitrary
complex number with $Re \lambda_k < 0$ (see below). The mass
squared parameter $\Delta^2$ (the above-mentioned mass gap) is
responsible for the scale of NP dynamics in the IR region in our
approach. The functions $f_{\lambda_k}(q^2)$ are, by definition,
dimensionless, regular at zero, and otherwise remaining arbitrary,
but preserving AF in the UV limit. And finally the number $k$ is a
positive integer, i.e., $k=0,1,2,3,....$ (see above). Evidently, a
real INP gluon form factor $d^{INP}(q^2)$, which now should depend
on the mass gap as well, i.e., $d^{INP}(q^2) \equiv d^{INP}(q^2,
\Delta^2)$, is a sum over all $d^{INP}_{\lambda_k}(q^2,
\Delta^2)$.

However, this is not the whole story yet. Since we are especially
interested in the deep IR structure of the full gluon propagator,
the arbitrary functions $f_{\lambda_k}(q^2)$ should be also
expanded around zero in the form of the Taylor series in powers of
$q^2$, i.e,

\begin{equation}
f_{\lambda_k}(q^2) = \sum_{m=0}^{[- \lambda_k] -(n/2)} {(q^2)^m
\over m!} f^{(m)}_{\lambda_k}(0) + \sum_{m=[-\lambda_k] - (n/2)
+1}^{\infty} {(q^2)^m \over m!} f^{(m)}_{\lambda_k}(0),
\end{equation}
where $[-\lambda_k]$ denotes its integer number and $n$ is a
number of the components in the Euclidean squared form $q^2$ (see
above). Also

\begin{equation}
f^{(m)}_{\lambda_k}(0) =  \Biggl( {d^m f_{\lambda_k}(q^2) \over
d(q^2)^m} \Biggr)_{q^2=0}.
\end{equation}
As a result, we will be left with the finite sum of power terms
with an exponent decreasing by unity starting from $- \lambda_k$.
All other remaining terms from the Taylor expansion (3.6),
starting from the term having already a PT IR singularity (the
second sum in Eq. (3.6)), should be shifted to the PT part of the
full gluon propagator in Eq. (3.2). So the INP part of the full
gluon form factor becomes

\begin{equation}
d^{INP}(q^2, \Delta^2)  = \sum_{\lambda_k}
d^{INP}_{\lambda_k}(q^2, \Delta^2)  =
\sum_{\lambda_k}(\Delta^2)^{-\lambda_k - 1}
 (q^2)^{\lambda_k}
\sum_{m=0}^{[- \lambda_k] -(n/2)} {(q^2)^m \over m!}
f^{(m)}_{\lambda_k}(0),
\end{equation}
while the piece which is to be shifted to the PT part (3.4) of the
full gluon propagator (3.2) can be shown to have only PT IR
singularities with respect to the gluon momentum, indeed
\cite{13}.

\section{4D QCD}

We are particulary interested in 4D QCD (i.e., $n=4$), which is a
realistic dynamical theory of strong interactions not only at the
fundamental quark-gluon level, but at the hadronic level as well.
Let us discuss the gluon propagator (3.2) in more detail for 4D
QCD, i.e., QCD itself. On account of the expansion (3.8) and Eq.
(2.1) at $n=4$ with the obvious identification $\lambda_k \equiv
\lambda$, its INP part becomes

\begin{equation}
d^{INP}(q^2, \Delta^2) = \sum_{k=0}^{\infty} d^{INP}_k(q^2,
\Delta^2)= \sum_{k=0}^{\infty} (\Delta^2)^{1 + k} (q^2)^{-2 -k}
\sum_{m=0}^{k} {(q^2)^m \over m!} f^{(m)}_k(0),
\end{equation}
and $f^{(0)}_k(0) \equiv f_k(0)$.
 Obviously, in this case the subscript "$\lambda_k$" should be
replaced by the subscript "$k$", since $\lambda_k =-2-k, \
k=0,1,2,3,...$. Thus $d^{INP}(q^2, \Delta^2)$ describes the true
(physical) NP vacuum of QCD, while $d^{INP}_k(q^2, \Delta^2)$
describe the auxiliary ones, and the former is an infinite sum of
the latter ones. The expansion (4.1) is obviously the Laurent
expansion in the inverse powers of the gluon momentum squared,
which every term ends at the simplest NP IR singularity
$(q^2)^{-2}$. The only physical quantity (apart from the mass gap,
of course) which can appear in this expansion is the coupling
constant squared in the corresponding powers. In QCD it is
dimensionless and is evidently included into the $f_k$ functions.
Let us note in advance that all the finite numerical factors and
constants (for example, the coupling constant) play no independent
role in the presence of a mass gap.

It is instructive to show explicitly expansions for a few first
different $d^{INP}_k(q^2, \Delta^2)$, namely

\begin{eqnarray}
d^{INP}_0(q^2, \Delta^2)&=& \Delta^2 f_0(0)(q^2)^{-2},
\nonumber\\
d^{INP}_1(q^2, \Delta^2)&=& (\Delta^2)^2 f_1(0)(q^2)^{-3} +
(\Delta^2)^2 f^{(1)}_1(0) (q^2)^{-2},  \nonumber\\
d^{INP}_2(q^2, \Delta^2)&=&(\Delta^2)^3 f_2(0)(q^2)^{-4} +
(\Delta^2)^3 f^{(1)}_2(0) (q^2)^{-3} + {1 \over 2} (\Delta^2)^3
f^{(2)}_2(0)(q^2)^{-2},
 \nonumber\\
d^{INP}_3(q^2, \Delta^2)&=& (\Delta^2)^4 f_3(0)(q^2)^{-5} +
(\Delta^2)^4 f^{(1)}_3(0)(q^2)^{-4} + {1 \over 2} (\Delta^2)^4
f^{(2)}_3(0)(q^2)^{-3} + {1 \over 6} (\Delta^2)^4
f^{(3)}_3(0)(q^2)^{-2}, \nonumber\\
\end{eqnarray}
and so on. Apparently, there is no way that such kind of an
infinite series (as it is present in expansion (4.1), on account
of the relations (4.2)) could be summed up into the finite
functions, i.e., functions which could be regular at zero. That is
why the above-mentioned smooth gluon propagator is, in general,
very unlikely to exist. Let also note in advance that the simplest
NP IR singularity $(q^2)^{-2}$ is present in each expansion, which
emphasizes its special and important role (see below).

The expansion (4.1), on account of the relations (4.2), can be
equivalently written down as follows:

\begin{equation}
d^{INP}(q^2, \Delta^2) = \sum_{k=0}^{\infty} (q^2)^{-2 -k}
\sum_{m=0}^{\infty} {1 \over m!} (\Delta^2)^{k+m+1}
f^{(m)}_{k+m}(0) = \sum_{k=0}^{\infty} (q^2)^{-2
-k}(\Delta^2)^{k+1} \sum_{m=0}^{\infty} {1 \over m!}
\varphi_{k,m}(0),
\end{equation}
where we use the relation

\begin{equation}
f^{(m)}_{k+m}(0) = (\Delta^2)^{-m} \varphi_{k,m}(0),
\end{equation}
which obviously follows from the relation (3.7), since all
$f^{(m)}_k(0)$ have the dimensions of the inverse mass squared in
powers of $m$, i.e., $[f^{(m)}_k(0)]= [\Delta^{-2}]^m  =
[\Delta^2]^{-m}$. Here $\varphi_{k,m}(0)$ are dimensionless
quantities, by definition. This expansion explicitly shows that
the coefficient at each NP IR singularity is an infinite series
itself. It also shows that we can analyze the IR properties of the
INP part of the full gluon form factor in terms of the mass gap
$\Delta^2$ and the dimensionless quantities $\varphi_{k,m}(0)$
only, which is very convenient from a technical point of view (see
below). This form of the Laurent expansion shows also clearly the
dynamical context of the INP part of the full gluon propagator.

The regularization of the NP IR singularities in QCD is determined
by the Laurent expansion (2.4) at $n=4$ as follows:

\begin{equation}
(q^2)^{- 2 - k } = { 1 \over \epsilon} a(k)[\delta^4(q)]^{(k)} +
f.t., \quad \epsilon \rightarrow 0^+,
\end{equation}
where $a(k)$ is a finite constant depending only on $k$ and
$[\delta^4(q)]^{(k)}$ represents the $k$th derivative of the
$\delta$-function (see Eq. (2.3)). We point out that after
introducing this expansion everywhere one can fix the number of
dimensions, indeed, i.e., put $D=n=4$ for QCD without any further
problems, since there will be no other severe IR singularities
with respect to $\epsilon$ as it goes to zero, but those
explicitly shown in this expansion. Thus, as it follows from the
Laurent expansion (4.5), any power-type NP IR singularity,
including the simplest one, scales as $1 /\epsilon$ as it goes to
zero. Just this plays a crucial role in the IR renormalization of
the theory within our approach.

\section{IR renormalization of a  mass gap}

 In the presence of such severe IR singularities
(4.5), all the "bare" parameters (dimensional or not, does not
matter) should, in principle, depend on $\epsilon$ as well, i.e.,
they become IR regularized. Let us thus introduce the following
relation

\begin{equation}
\Delta^2 = X_{\Delta}(\epsilon) \bar \Delta^2, \quad \epsilon
\rightarrow 0^+,
\end{equation}
where $X_{\Delta}(\epsilon)$ is the corresponding IR
multiplicative renormalization (IRMR) constant. Here and below,
the quantities with an overbar are, by definition, IR
renormalized, i.e., they exist as $\epsilon$ goes to zero.
However, in the above-mentioned paper \cite{13} it has been proven
that neither the QCD coupling constant squared nor the gauge
fixing parameter are to be IR renormalized, i.e., they are IR
finite from the very beginning. As was mentioned above, they can
appear only in $f^{(m)}_k(0)$ quantities, i.e., in
$\varphi_{k,m}(0)$ ones. So these quantities also are IR finite
from the very beginning, which means that we can put
$\varphi_{k,m}(0) \equiv \bar \varphi_{k,m}(0)$. This is so
indeed, since the rest in these quantities is simply the product
of the numerical factors like $\pi$'s in negative powers,
eigenvalues of the color group generators (we are not considering
the numbers of different colors and flavors as free parameters of
the theory), etc.

We already know that all the NP IR singularities, which can appear
in the full gluon propagator scale as $1/ \epsilon$ with respect
to $\epsilon$ (see Eq. (4.5)).  So let us introduce the so-called
IR convergence condition as follows:

\begin{equation}
X_{\Delta}^{1+k}(\epsilon) = \epsilon \tilde{A}_k, \quad \epsilon
\rightarrow 0^+,
\end{equation}
where we put $\tilde{A}_k =  \bar A_k / \sum_{m=0}^{\infty} (1 /
m!) \bar \varphi_{k,m}(0)$, for convenience. The cancellation of
the NP IR singularities with respect to $\epsilon$ then will be
guaranteed in Eq. (4.3), and one obtains the finite (nonzero)
result in the $\epsilon \rightarrow 0^+$ limit for every
$k=0,1,2,3,...$. Here $\bar A_k$ and  $\tilde{A}_k$ are some
arbitrary, but finite constants, not depending on $\epsilon$ as it
goes to zero.

 It makes sense to emphasize now that this IR convergence
condition should be valid at any $k$, in particular at $k=0$, so
it follows $X_{\Delta}(\epsilon) = \epsilon \tilde{A}_0, \quad
\epsilon \rightarrow 0^+$, which means that in this case one is
able to establish an explicit solution for the mass gap's IRMR
constant. Thus the mass gap is IR renormalized as follows:

\begin{equation}
\Delta^2 = \epsilon \bar \Delta^2, \quad \epsilon \rightarrow 0^+,
\end{equation}
where we include an arbitrary but finite constant $\tilde{A}_0$
into the IR renormalized mass gap $\bar \Delta^2$, and retaining
the same notation, for simplicity. This means that in what follows
we can put it to unity, not losing generality, i.e., $\tilde{A}_0
=1$.

 It is instructive to rewrite the expansion (4.3) as follows:

\begin{equation}
d^{INP}(q^2, \Delta^2) = \Delta^2 (q^2)^{-2} \sum_{m=0}^{\infty}
{1 \over m!} \bar \varphi_{0,m}(0) + \sum_{k=1}^{\infty} (q^2)^{-2
-k}(\Delta^2)^{k+1} \sum_{m=0}^{\infty} {1 \over m!} \bar
\varphi_{k,m}(0).
\end{equation}
Taking into account now the above-described scaling with respect
to $\epsilon$, the INP part of the full gluon form factor
effectively becomes

\begin{equation}
d^{INP}(q^2, \Delta^2)= \Delta^2 (q^2)^{-2} \sum_{m=0}^{\infty} {1
\over m!} \bar \varphi_{0,m}(0),
\end{equation}
where finite dimensionless quantities $\bar \varphi_{0,m}(0)$
contain the IR finite coupling constant squared (in different
powers, of course, so the whole expansion in $m$ includes all
powers in the coupling constant squared). It also contains the
different combinations of the finite numerical factors only.
Evidently, no other terms, explicitly shown in the expansion (5.4)
as the second sum, will survive in the $\epsilon \rightarrow 0^+$
limit. They become terms of the order of $\epsilon$, at least, in
this limit (they start from $(\Delta^2)^2 \sim \epsilon^2$, while
$(q^2)^{-2-k}$ always scales as $1 / \epsilon$). In other words,
in the Laurent expansion (5.4) only the terms which contain the
simplest NP IR singularity with respect to the gluon momentum
$(q^2)^{-2}$ will survive as $\epsilon \rightarrow 0^+$. In this
case all terms (apart from the simplest NP IR singularity) will be
additionally suppressed in positive powers of $\epsilon$. The IR
renormalization of the mass gap $\Delta^2$ in accordance with Eq.
(5.3) cancels $1 / \epsilon$, which comes from the regularization
of $(q^2)^{-2}$. The so-called "f.t." terms in the Laurent
expansion (4.5) become terms of the order of $\epsilon$, at least,
so here and everywhere they vanish in the $\epsilon \rightarrow
0^+$ limit.

\section{ZMME quantum model of the QCD ground state}

The true QCD ground state is a very complicated confining medium,
containing many types of gluon field configurations, components,
ingredients and objects of different nature. Its dynamical and
topological complexity means that its structure can be organized
at both the quantum and classical levels \cite{1,17}. Our quantum,
dynamical model of the true QCD true vacuum is based on the
existence and the importance of such kind of the NP excitations
and fluctuations of gluon field configurations which are due to
the self-interaction of massless gluons only, without explicitly
involving some extra degrees of freedom. They are to be summarized
(accumulated) into the INP part of the full gluon propagator, and
are to be effectively correctly described by its strongly singular
structure in the deep IR domain (for simplicity, we will call them
as singular gluon field configurations). At this stage, it is
difficult to identify actually which type of gauge field
configurations can be behind the singular gluon field
configurations in the QCD ground state (i.e., to identify relevant
field configurations: chromomagnetic, self-dual, stochastic,
etc.). However, if these gauge field configurations can be
absorbed into the gluon propagator (i.e., if they can be
considered as solutions to the corresponding SD equation), then
its severe IR singular behavior is a common feature for all of
them. Being thus a general phenomenon, the existence and the
importance of quantum excitations and fluctuations of the severely
IR degrees of freedom inevitably lead to the ZMME effect in the
QCD ground state. That is why we call our model of the QCD ground
state as the ZMME quantum model, or simply zero modes enhancement
(ZME, since we work always in the momentum space). For preliminary
investigation of this model see our papers \cite{18,19} and
references therein.

Our approach to the QCD true ground state, based on the ZMME
phenomenon there, in terms of the gluon propagator, can be
analytically formulated as follows:

\begin{equation}
D_{\mu\nu}(q) = D^{INP}_{\mu\nu}(q,\Delta) + D^{PT}_{\mu\nu}(q),
\end{equation}
where the INP part of the full gluon propagator before the IR
renormalization is

\begin{equation}
D^{INP}_{\mu\nu}(q,\Delta) = i T_{\mu\nu}(q) d^{INP}(q^2,
\Delta^2) = i T_{\mu\nu}(q) \times \Bigl[ \Delta^2 \bar A_0
(q^2)^{-2} + \sum_{k=1}^{\infty} (\Delta^2)^{1 + k} a_k (q^2)^{-2
-k} \Bigr],
\end{equation}
where $a_k =(\bar A_k / \tilde{A}_k)$ and we used Eq. (5.4), on
account of the relation (5.2) for the coefficients. In fact, Eq.
(6.2) is already partially IR renormalized with respect to the
coefficients, but not with respect to the mass gap $\Delta^2$ and
the NP IR singularities $(q^2)^{-2 -k}$. After the IR
renormalization it effectively becomes

\begin{equation}
D^{INP}_{\mu\nu}(q, \Delta) = i T_{\mu\nu}(q) d^{INP}(q^2,
\Delta^2) = i T_{\mu\nu}(q) \Delta^2 (q^2)^{-2},
\end{equation}
since only the first term in Eq.(6.2) survives in the $\epsilon
\rightarrow 0^+$ limit, as was explained above. An arbitrary but
finite constant $\bar A_0$ has been included into the mass gap
with retaining the same notation, for convenience. The PT part of
the full gluon propagator in any case remains undetermined (see
remaks below).

\subsection{Confinement criterion for gluons}

It is worth discussing the properties of the obtained solution for
the full gluon propagator in more detail. It is already clear that
by the IR renormalization of the mass gap only, we can render the
full gluon propagator IR finite from the very beginning, t.e., to
put $D(q) \equiv \bar D(q) =  \bar D^{INP}(q) + \bar D^{PT}(q)$
(this has been rigorously proven in Ref. \cite{13}). In principle,
two different cases should be considered due to the distribution
nature of the simplest NP IR singularity$(q^2)^{-2}$, which
saturates its INP part in Eq. (6.3).

I. If there is the explicit integration over the gluon momentum,
then from Eq. (6.3) it follows

\begin{equation}
\bar D^{INP}(q) \equiv \bar D^{INP}_{\mu\nu}(q, \bar \Delta)=
iT_{\mu\nu}(q) \bar \Delta^2 \pi^2 \delta^4(q),
\end{equation}
i.e., in this case we have to replace the NP IR singularity
$(q^2)^{-2}$ in Eq. (6.3) by its $\delta$-type regularization
(4.5) at $k=0$. We also always should take into account the
relation (5.3) for the IR renormalization of the mass gap in order
to express all relations in terms of the IR renormalized
quantities. The $\delta$-type regularization is valid even for the
multi-loop skeleton diagrams, where the number of independent
loops is equal to the number of the gluon propagators. In the
multi-loop skeleton diagrams, where these numbers do not coincide
(for example, in the diagrams containing three or four-gluon
proper vertices), the general regularization (4.5) should be used
(i.e., derivatives of the $\delta$-functions), and not the product
of the $\delta$-functions at the same point, which has no
mathematical meaning in the DT sense \cite{14}.

II. If there is no explicit integration over the gluon momentum,
then the full gluon propagator is reduced to $D(q) \equiv \bar
D_{\mu\nu}(q) = \bar D^{PT}_{\mu\nu}(q)$ in the $\epsilon
\rightarrow 0^+$ limit, since in this case the function
$(q^2)^{-2}$ in Eq. (6.3) cannot be treated as the distribution.
Only the relation (5.3) again comes out into the play.
 So the INP part of the full gluon propagator (6.3) in this
case disappears as $\epsilon $ as $\epsilon \rightarrow 0^+$,
namely

\begin{equation}
\bar D^{INP}_{\mu\nu}(q, \bar \Delta^2) \sim \epsilon, \quad
\epsilon \rightarrow 0^+.
\end{equation}
This means that any amplitude (more precisely its INP part, see
below) for any number of soft-gluon emissions (no integration over
their momenta) will vanish in the IR limit in our picture. In
other words, there are no transverse gluons in the IR, i.e., at
large distances (small momenta) there is no possibility to observe
gluons experimentally as free particles. So this behavior can be
treated as the gluon confinement criterion (see also Ref.
\cite{12}), and it supports the consistency of the exact solution
(6.3) for the INP part of the full gluon propagator. Evidently,
this behavior does not explicitly depend on the gauge choice in
the full gluon propagator, i.e., it is a manifestly
gauge-invariant as it should be, in principle. Concluding this
Sect., it is worth underlining that the gluon confinement
criterion (6.5) is valid in the general case as well, i.e., before
explicitly showing that the QCD coupling constant squared and the
gauge fixing parameter are IR finite \cite{13}.

subsection{INP phase in QCD}

For the sake of self-consistency and transparency of our approach
to low-energy QCD, it is convenient to discuss in more detail what
we mean by the INP phase in QCD. The INP part of the full gluon
propagator is, in general, given in Eqs. (6.4) and (6.5) for the
above-discussed two different cases. Let us remind that within our
approach all the severe IR singularities of the dynamical origin
possible in QCD are to be incorporated into the full gluon
propagator and are to be effectively correctly described by its
INP part. So all other QCD Green's functions can be considered as
regular functions with respect to all the gluon momenta involved.
If, nevertheless, they might be singular, then it would require a
completely different investigation, anyway.

 In principle, all other fundamental quantities in QCD could be
formally decomposed similar to the decomposition (6.1) for the
full gluon propagator. Evidently, this should be done for
quantities, which explicitly depend on the gluon momenta, i.e.,
proper vertices. The decomposition does not make any sense for the
coupling constant, quark and ghost propagators, since they do not
depend on the gluon momentum. This is also true for the quark- and
ghost-gluon proper vertices, since they depend on the quark and
ghost momenta, which are completely independent from the gluon
momentum, playing the role of the momentum transfer in these
vertices. So the only proper vertices which makes sense to
decompose are the pure gluon vertices, since they crucially depend
on all the gluon momenta involved.

However, we define the INP phase in QCD in more general terms,
which includes the decomposition of the full gluon propagator only
as follows:

(i) It is always transverse, i.e., it depends only on physical
degrees of freedom of gauge bosons.

(ii) Before the IR renormalization, the presence of the NP IR
singularities $(q^2)^{-2-k}, \ k=0,1,2,3,...$ is only possible.

(iii). After the IR renormalization, the INP part of the full
gluon propagator is fully saturated by the simplest NP IR
singularity and all other NP IR singularities will be additionally
suppressed in the $\epsilon \rightarrow 0^+$ limit.

(iv). There is an inevitable dependence on the mass gap
$\Delta^2$, so that when it formally goes to zero, then the INP
phase vanishes, while the PT phase survives.

This definition implies that the INP part of any multi-loop
skeleton diagram in QCD should contain only the INP parts of all
the corresponding gluon propagators. At the same time, the PT part
of any multi-loop skeleton diagram always remains of arbitrary
gauge. It may even contain the terms, where the NP IR
singularities are present along with the PT IR ones as well (the
so-called general PT term). Let us also remind, that at the level
of a single full gluon propagator, its PT part is defined as to be
of arbitrary gauge and is as much singular as $1 /q^2$. The
difference between the NP IR singularities $(q^2)^{-2-k}$ and the
PT IR singularity $(q^2)^{-1}$ is that the latter is not defined
by its own Laurent expansion (4.5) that is dimensionally
regularized like the former ones. That is why it does not require
the IR renormalization program itself.

\section{Discussion}

Evidently, the ZMME mechanism for quark confinement is nothing but
the well forgotten IR slavery (IRS) one, which can be equivalently
referred to as a strong coupling regime \cite{1}. Indeed, at the
very beginning of QCD it was expressed a general idea
\cite{5,6,7,8,9,10,11,12} that the quantum excitations of the IR
degrees of freedom, because of self-interaction of massless gluons
in the QCD vacuum, made it only possible to understand
confinement, DCSB and other NP effects. In other words, the
importance of the deep IR structure of the true QCD vacuum has
been emphasized as well as its relevance to quark confinement,
DCSB, etc., and the other way around. This development was stopped
by the wide-spread wrong opinion that the severe IR singularities
cannot be put under control. We have explicitly shown that the
correct mathematical theory of quantum YM physical theory is the
theory of distributions (theory of generalized functions)
\cite{14}, complemented by the DR method \cite{15}. They provide a
correct treatment of these severe IR singularities without any
problems. Thus we come back to the old idea but on a new basis
that is why it becomes new ("new is well forgotten old"). In other
words, we only put the IRS mechanism of quark confinement on a
firm mathematical ground provided by DT.

There are a lot of direct and indirect evidences in favor of this
behavior, Eq. (6.3) \cite{13}. Let us note only that as a possible
IR asymptotics of the full gluon propagator in different covariant
and non-covariant gauges, it has been already successfully
investigated in Refs. \cite{20,21,22,23,24,25}. Following the
pioneering paper of Pagels \cite{12}, we always used it in our
preliminary investigation of the quark confinement problem (see
Refs. \cite{18,19} and references therein). It is worth
emphasizing, however, that Eq. (6.3) is an exact result, and
expresses not only the deep IR asymptotics of the full gluon
propagator. Just this behavior (6.3) in the continuous theory
provides the linear rising potential (indicative of confinement)
between heavy quarks "seen" by lattice simulations \cite{3}.

\section{Conclusions}

Emphasizing the highly nontrivial structure of the true QCD ground
state in the deep IR region, one can conclude:

1). The self-interaction of massless gluons is only responsible
for the ZMME effect in the true QCD vacuum, which in its turn, is
to be taken into account by the deep IR structure of the full
gluon propagator.

2). The full gluon propagator thus is inevitably more singular in
the IR than its free counterpart.

3). This requires the existence of a mass gap, which is
responsible for the NP dynamics in the QCD vacuum. It appears on
dynamical ground due to the self-interaction of massless gluons
only.

4). We decompose algebraically (i.e., exactly) the full gluon
propagator as a sum of its INP and PT parts. We additionally
distinguish between them dynamically by the different character of
the IR singularities in each part.

5). We have exactly established the general IR structure of the
full gluon propagator in QCD as an infinite sum over all possible
NP IR singularities (see the expansion (4.1) or equivalently
(4.3)).

6). The next step is to regularize them correctly, i.e., to use
the Laurent expansion (4.5) that is dimensionally regularized.

7). We emphasize once more that the IR renormalization program is
based on an important observation that the NP IR singularities
$(q^2)^{-2-k}$, being distributions, always scale as $1 /
\epsilon$, not depending on the power of the singularity $k$,
i.e., $(q^2)^{-2-k} \sim \ 1/ \epsilon$. It is easy to understand
that otherwise none of the IR renormalization program in the INP
phase of QCD  and QCD itself would be possible.

8). The IR renormalization of the initial mass gap (Eq. (5.3)) is
only needed in order to fix uniquely and exactly the IR structure
of the full gluon propagator in QCD. It is saturated by the
simplest NP IR singularity, the famous $(q^2)^{-2}$, Eq. (6.3).
The mass gap gains contributions from all orders of PT in the QCD
coupling constant squared, which remains IR finite.

9). On this basis, we have formulated the ZMME model of the true
QCD ground state. Due to the distribution nature of the NP IR
singularities, two different types of the IR renormalization of
the INP part of the full gluon propagator are required,
preserving, nevertheless, its IR finiteness.

10). This makes it possible to establish the gluon confinement
criterion in a manifestly gauge-invariant way in Eq. (6.5).

11). In the same way, we define exactly the INP phase in QCD at
the fundamental gluon level. The corresponding decomposition of
the full gluon propagator is only needed in order to firmly
control the IR region in QCD within our approach (only it contains
explicitly the mass gap).

 Our general conclusions are:

I. The NP structure of the true QCD ground state is to be
described by the IR structure of the full gluon propagator.

II. In its turn, it is an infinite sum over all possible NP IR
singularities. Due to their distribution nature, any solution to
the gluon SD equation has to be always present in the form of the
corresponding Laurent expansion. It makes it possible to control
both, the whole expansion as well as its each term, by the correct
use of DT.

III. The mass gap responsible for the NP dynamics in the true QCD
ground state is required. This is important, since there is none
in the QCD Lagrangian.

 IV. Complemented by the DR
method, DT puts the treatment of the NP IR singularities on a firm
mathematical ground. So there is no place for theoretical
uncertainties. The wide-spread opinion that they cannot be
controlled is not justified.

V. This makes it possible to fix uniquely the IR structure of the
full gluon propagator in QCD, not solving directly the
corresponding SD equation itself. Thus we have exactly established
the interaction between quarks (concerning its pure gluon (i.e.,
nonlinear) contribution up to its unimportant PT part).

VI. This somehow astonishingly radical result has been achieved at
the expense of the PT part of the full gluon propagator. It
remains of arbitrary covariant gauge and its functional dependence
cannot be determined.

However, let us note in advance that our theory (which we call INP
QCD) will be additionally defined by the subtraction of all types
of the PT contributions in order to completely decouple it from
QCD as a whole (the first step in these subtractions has been
already done in Eq. (3.1)). To discuss this point and many other
interesting features of our approach in detail \cite{13} is, of
course, beyond the scope of this Letter.

A financial support from HAS-JINR Scientific Collaboration Fund
(P. Levai) and Hungarian Scientific Fund OTKA T30171 (K. Toth) is
to be acknowledged.

\end{document}